\documentclass[aps,twocolumn]{revtex4}
\usepackage[dvips]{graphicx}
\begin{document}

\title{ {\tt {\small \begin{flushright}
MAN/HEP/2009/15, April 2009
\end{flushright} } }
Supersymmetric Lepton Flavour Violation in Low-Scale Seesaw Models}

\author{Amon Ilakovac$^{\,a}$ and Apostolos Pilaftsis$^{\,b}$\vspace{2mm}}

\affiliation{${}^a$University of Zagreb, Department of Physics,
Bijeni\v cka cesta 32, P.O. Box 331, Zagreb, Croatia\vspace{1mm}\\
${}^b$School of Physics and Astronomy, University of
Manchester, Manchester M13 9PL, United Kingdom}

\begin{abstract}
\noindent
We study  a new supersymmetric mechanism for  lepton flavour violation
in  $\mu$ and  $\tau$ decays  and $\mu  \to e$  conversion  in nuclei,
within a  minimal extension  of the MSSM  with low-mass  heavy singlet
neutrinos and sneutrinos.  We find  that the decays $\mu \to e\gamma$,
$\tau  \to e\gamma$  and $\tau  \to  \mu\gamma$ are  forbidden in  the
supersymmetric limit  of the theory, whereas other  processes, such as
$\mu \to  eee$, $\mu  \to e$ conversion,  $\tau \to eee$  and $\tau\to
e\mu\mu$, are allowed and  can be dramatically enhanced several orders
of magnitude above the  observable level by potentially large neutrino
Yukawa coupling effects.   The profound implications of supersymmetric
lepton  flavour  violation  for  present and  future  experiments  are
discussed.

\medskip

\end{abstract}

\maketitle 

One of  the best theoretically  motivated scenarios of new  physics is
the Minimal Supersymmetric Standard Model (MSSM), softly broken at the
TeV   scale.    Its   main    virtues   are   that   it   provides   a
quantum-mechanically stable solution  to the so-called gauge hierarchy
problem, predicts gauge-coupling  unification more accurately than the
Standard  Model (SM)  does  and  offers a  hopeful  perspective for  a
consistent quantization  of gravity  by means of  supergravity (SUGRA)
and superstrings~\cite{Review}.

Nevertheless, the low-energy  sector of the MSSM needs  be extended in
order to  accommodate the  low-energy neutrino oscillation  data.  One
popular  extension  is  the   one  that  realizes  the  famous  seesaw
mechanism~\cite{seesaw},  where   the  smallness  of   the  observable
neutrinos   is  counter-balanced   by  the   presence   of  ultraheavy
right-handed neutrinos  $N_{1,2,3}$ with Majorana masses  that are two
to four orders  of magnitude below the grand  unification theory (GUT)
scale  $M_{\rm GUT}  \sim 10^{16}$~GeV.   The  leptonic superpotential
part for this extension is
\begin{equation}
  \label{Wpot}
W_{\rm lepton}\ =\  \widehat{E}^C {\bf h}_e \widehat{H}_d
\widehat{L}\: +\: \widehat{N}^C {\bf h}_\nu \widehat{L} \widehat{H}_u\:
+\: \widehat{N}^C {\bf m}_M \widehat{N}^C\; ,
\end{equation}
where    $\widehat{H}_{u,d}$,    $\widehat{L}$,   $\widehat{E}$    and
$\widehat{N}^C$  denote the two  Higgs-doublet superfields,  the three
left-  and  right-handed  charged-lepton  superfields  and  the  three
right-handed neutrino superfields,  respectively. Note that the Yukawa
couplings~${\bf  h}_{e,\nu}$ and  the  Majorana mass  parameters~${\bf
m}_M$  are $3\times  3$ complex  matrices. In  a minimal  SUGRA seesaw
model, lepton  flavour violation (LFV),  such as $\mu\to  e\gamma$ and
$\mu  \to  eee$,  originates from  off-diagonal  renormalization-group
effects induced by the  neutrino Yukawa couplings~${\bf h}_\nu$ on the
soft    supersymmetry     (SUSY)    breaking    mass    matrices~${\bf
\widetilde{M}}^2_{L,E}$  and the  trilinear couplings~${\bf  h}_e {\bf
A}_e$~\cite{BM,HMTY}.  However,  if the soft  SUSY-breaking parameters
${\bf \widetilde{M}}^2_{L,E}$ and ${\bf A}_e$ were flavour diagonal or
proportional  to the  3-by-3 identity  matrix  {\bf 1}  at ${\bf  m}_M
\approx  m_N {\bf 1}$,  with $m_N  \stackrel{>}{{}_\sim} 10^{12}$~GeV,
all low-energy charged LFV  phenomena would be extremely suppressed by
factors  $m_\nu/m_N$~\cite{CL},  where $m_{\nu}  \stackrel{<}{{}_\sim}
0.1$~eV is the light-neutrino mass scale.

In  this Letter  we study  a new  supersymmetric mechanism  for lepton
flavour  violation  (SLFV)  which  becomes  dramatically  enhanced  in
low-scale seesaw  extensions of  the MSSM.  As  we will show,  the new
important  feature  of  SLFV  is  that  it  does  not  vanish  in  the
supersymmetric  limit  of  the  theory,  giving  rise  to  distinctive
predictions for charged LFV in present and future experiments, such as
MEG~\cite{MEG} and PRISM~\cite{PRISM}.

In low-scale seesaw models of interest here~\cite{WW,MV,BGL,AZPC}, the
smallness of  the light neutrino  masses is accounted for  by natural,
quantum-mechanically  stable  cancellations  due  to the  presence  of
approximate lepton  flavour symmetries~\cite{AZPC,APRLtau}, whilst the
Majorana  mass   scale  $m_N$  can   be  as  low  as   100~GeV.   Most
interestingly, in  these models LFV transitions from  a charged lepton
$l=e,\,\mu,\,\tau$ to  another $l'\neq l$ are  generically enhanced by
the ratios~\cite{KPS,IP,DV}
\begin{equation}
  \label{Omega}
{\bf  \Omega}_{ll'}\ =\  \frac{v^2_u}{2  m^2_N}\ ({\bf  h}^\dagger_\nu
{\bf h}_\nu)_{ll'}
\end{equation}
and are not constrained by  the usual seesaw factor $m_\nu/m_N$, where
$v_u/\sqrt{2}  \equiv \langle  H_u\rangle$ is  the  vacuum expectation
value (VEV) of the Higgs doublet $H_u$, with $\tan\beta \equiv \langle
H_u\rangle/\langle  H_d\rangle$.   Here  we  will set  limits  on  the
off-diagonal     entries    of     ${\bf     \Omega}_{e\mu}$,    ${\bf
\Omega}_{\mu\tau}$  and ${\bf \Omega}_{e\tau}$  that are  derived from
the non-observation of  LFV in $\mu$ and $\tau$  decays and of $\mu\to
e$ conversion in nuclei~\cite{OmegaDiag}.

To be able to understand  the profound implications of SLFV, we assume
that  the singlet  neutrino sector  of the  low-scale seesaw  model is
exactly supersymmetric.   This assumption is a  good approximation, as
long  as $m_N  \gg M_{\rm  SUSY}$, where  $M_{\rm  SUSY} =$~0.1--1~TeV
denotes  a  typical  soft  SUSY-breaking  mass for  the  U(1)$_Y$  and
SU(2)$_L$  gauginos, $\widetilde{B}$ and  $\widetilde{W}_{1,2,3}$, and
for  the  left-handed  sneutrinos, $\tilde{\nu}_{e,\mu,\tau}$.  As  an
illustrative scenario, we consider that  $m_N$ is much larger than the
superpotential $\widehat{H}_u\widehat{H}_d$-mixing parameter $\mu$ and
that  ${\bf  \widetilde{M}}^2_{L,E}$   and  ${\bf  A}_e$  are  flavour
conserving, e.g.~proportional to {\bf 1} at the energy-scale $m_N$.
\begin{figure}[t]
 \centering
  \includegraphics[clip,width=0.26\textwidth,height=0.36\textheight,
     angle=90]{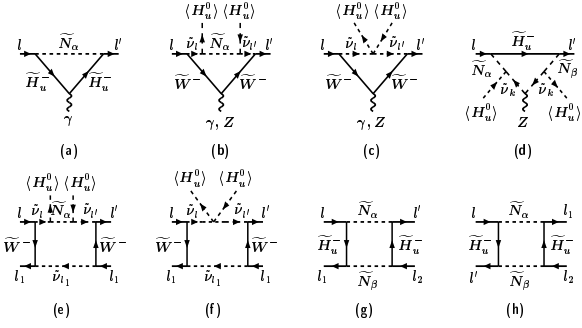} 
\caption{Feynman  graphs giving rise  to leading  SLFV effects  in the
lowest   order  of   an  expansion   in  $\langle   H^0_u\rangle$  and
$m^{-1}_N$. Not  shown are diagrams  obtained by replacing  the tilted
SUSY        states       $\widetilde{H}^-_u$,       $\widetilde{W}^-$,
$\widetilde{N}_{\alpha}$  and   $\tilde{\nu}_l$  with  their  untilted
counterparts.}\label{f1}
\end{figure}

Within the above simplified  but realistic framework, we may calculate
the leading effects of SLFV in  the lowest order of a series expansion
of $v_u$  and $m_N^{-1}$.  We  ignore charged Higgs effects  which are
subdominant in such an expansion.  Detailed analytic results including
this and  other subleading contributions  will be given in  a separate
communication~\cite{IPslfv}.   The Feynman  graphs that  contribute to
$\gamma l'l$-  and $Zl'l$-couplings and box diagrams  to leading order
in the SU(2)$_L$ gauge coupling $g_w$ and the neutrino Yukawa coupling
${\bf  h}_\nu$ are  shown in  Fig.~\ref{f1}. The  pertinent transition
amplitudes may be cast into the form:
\begin{eqnarray}
  \label{Trans}
{\cal T}^{\gamma l'l}_\mu \!&=&\! \frac{e\, \alpha_w}{8\pi M^2_W}\
\bar{l}' \Big( F_\gamma^{l'l}\, q^2 \gamma_\mu P_L + G^{l'l}_\gamma\, 
i\sigma_{\mu\nu} q^\nu m_l P_R \Big) l\;,\nonumber\\
{\cal T}^{Z l'l}_\mu \!&=&\!  \frac{g_w\, \alpha_w}{8\pi \cos\theta_w}\
F^{l'l}_Z\, \bar{l}' \gamma_\mu P_L l\; ,\\ 
{\cal T}^{l'l_1l_2}_l \!\!&=&\!  -\frac{\alpha^2_w}{4 M^2_W}\;
F^{ll'l_1l_2}_{\rm box}\, \bar{l}'\gamma_\mu P_L l\;
\bar{l}_1\gamma^\mu P_L l_2 \; ,\nonumber
\end{eqnarray}
where  $P_{L(R)}   =  \frac{1}{2}\,[1-\!(+)\,\gamma_5]$,  $\alpha_w  =
g^2_w/(4\pi)$, $e$  is the  electromagnetic coupling constant,  $M_W =
g_w \sqrt{v^2_u  +v^2_d}/2$ is the  $W$-boson mass, $\theta_w$  is the
weak mixing angle  and $q = p_{l'} - p_l$ is  the photon momentum.  In
addition,   the    formfactors   $F^{l'l}_\gamma$,   $G^{l'l}_\gamma$,
$F^{l'l}_Z$ and  $F^{ll'l_1l_2}_{\rm box}$ receive  contributions from
both the  heavy neutrinos $N_{1,2,3}$ and  the right-handed sneutrinos
$\widetilde{N}_{1,2,3}$.  In the Feynman gauge~\cite{Gauge}, these are
individually given by
\begin{eqnarray}
  \label{Fgamma}
(F^{l'l}_\gamma)^N \!&=&\! \frac{{\bf \Omega}_{l'l}}{6\,s^2_\beta}\, 
  \ln \frac{m^2_N}{M^2_W}\; ,\nonumber\\
(F^{l'l}_\gamma)^{\widetilde{N}} \!&=&\!  \frac{{\bf
  \Omega}_{l'l}}{3\,s^2_\beta}\, 
  \ln \frac{m^2_N}{\widetilde{m}^2_h}\; ,\\
  \label{Ggamma}
(G^{l'l}_\gamma)^N \!&=&\! -\,{\bf
  \Omega}_{l'l}\, \bigg(\, \frac{1}{6\,s^2_\beta}\: +\: 
                                        \frac{5}{6}\, \bigg)\;,\nonumber\\
(G^{l'l}_\gamma)^{\widetilde{N}} \!& = &\!  
{\bf   \Omega}_{l'l}\, \Bigg(\, \frac{1}{6\,s^2_\beta} \: +\:
  f\,\bigg)
\; ,\\
  \label{FZ}
(F^{l'l}_Z)^N \!&=&\! -\, \frac{3\, {\bf \Omega}_{l'l}}{2}\,
  \ln \frac{m^2_N}{M^2_W}\ -\ \frac{({\bf \Omega}^2)_{l'l}}{2\,s^2_\beta} \,
  \frac{m^2_N}{M^2_W}\; ,\nonumber\\
(F^{l'l}_Z)^{\widetilde{N}} \!&=&\! \frac{{\bf \Omega}_{l'l}}{2} \,
  \ln \frac{m^2_N}{\widetilde{m}^2_1}\ +\ \frac{({\bf
  \Omega}^2)_{l'l}}{4\,s^2_\beta} \,
  \frac{m^2_N}{M^2_W}\; \ln\frac{m^2_N}{\widetilde{m}^2_2}\; ,\qquad\\
  \label{Fbox}
(F^{ll'l_1l_2}_{\rm  box})^N \!\!&=&\! -\,{\bf
  \Omega}_{l'l}\,\delta_{l_1l_2} -  {\bf
  \Omega}_{l_1l}\,\delta_{l'l_2}\nonumber\\ 
\!&&\! +\: \frac{1}{4\,s^4_\beta}\, 
\Big( {\bf \Omega}_{l'l}\,{\bf \Omega}_{l_1l_2} + {\bf
\Omega}_{l_1l}\,{\bf \Omega}_{l'l_2}\,\Big)\,\frac{m^2_N}{M^2_W}\;,\nonumber\\
(F^{ll'l_1l_2}_{\rm  box})^{\widetilde{N}} \!\!&=&\!  
-\,\frac{M^2_W}{\widetilde{m}^2}
\Big({\bf \Omega}_{l'l}\,\delta_{l_1l_2} + {\bf
  \Omega}_{l_1l}\,\delta_{l'l_2}\Big)\nonumber\\ 
\!&&\! +\: \frac{1}{4\,s^4_\beta}\, 
  \Big( {\bf \Omega}_{l'l}\,{\bf \Omega}_{l_1l_2} + {\bf
  \Omega}_{l_1l}\,{\bf \Omega}_{l'l_2}\,\Big)\, \frac{m^2_N}{M^2_W}\; .
\end{eqnarray}
In the above, it is $s_\beta \equiv \sin\beta$, $\widetilde{m}^2_{W} =
{\rm max}\,  (M^2_{\widetilde{W}}, g^2_w v^2_u/2)$, $\widetilde{m}^2_h
= {\rm  max}\, (\mu^2, g^2_w v^2_u/2)$,  $\widetilde{m}^2_{1,2} = {\rm
  max}\,         (\widetilde{m}^2_{W,h},M^2_{\tilde{\nu}})$        and
$\widetilde{m}^2      =      {\rm     max}\,      (2M^2_{\tilde{\nu}},
\widetilde{m}^2_{W})$, where  $M_{\tilde{\nu}}$ is a  common soft SUSY
mass for  $\tilde{\nu}_{e,\mu,\tau}$. Moreover, the  loop function $f$
given  in~(\ref{Ggamma}) is  a  lengthy expression  involving all  the
above parameters~\cite{IPslfv}.   In the  SUSY limit which  requires a
vanishing  $\mu$-parameter and $\tan\beta  = 1$~\cite{HK,CommentSUSY},
$\widetilde{m}^2_{W,h}$, $\widetilde{m}^2_{1,2}$ and $\widetilde{m}^2$
all tend to~$M^2_W=g^2_wv^2_u/2$ and $f \to 5/6$.

It  is now  important to  notice that  the photonic  dipole formfactor
$G^{l'l}_\gamma$ vanishes  in the SUSY  limit when the  heavy neutrino
and      sneutrino     contributions      given     in~(\ref{Ggamma}),
$(G^{l'l}_\gamma)^N$ and $(G^{l'l}_\gamma)^{\widetilde{N}}$, are added
together.    This    result   is    a   direct   consequence    of   a
non-renormalization theorem  of SUSY~\cite{FR}.  Thus, if  SUSY is the
dominant  source for LFV  in nature,  photonic charged  lepton decays,
such  as $\mu  \to e\gamma$,  $\tau \to  e \gamma$  and $\tau  \to \mu
\gamma$, become  practically forbidden transitions.   In reality, SUSY
is softly  broken and  these decay rates  will strongly depend  on the
details  of the  soft SUSY-breaking  sector.  It  is  remarkable here,
however, that even a  flavour conserving soft SUSY-breaking sector for
the low-scale seesaw models under study can cause sizeable LFV.

Another  important observation  pertains  the actual  strength of  the
neutrino   Yukawa  couplings   ${\bf  h}_\nu$.    If   $|{\bf  h}_\nu|
\stackrel{>}{{}_\sim} g_w  \approx 0.65$,  then terms of  order $({\bf
\Omega}_{l'l})^2  \propto ({\bf  h}^\dagger_\nu  {\bf h}_\nu)^2_{l'l}$
will  dominate the  $Z$-boson and  box-mediated  transition amplitudes
given in~(\ref{Trans})~\cite{IP}.  Therefore, the central goal of this
study  is to  identify the  key phenomenological  features  that would
enable  one  to distinguish  whether  SLFV  originates  from small  or
potentially large neutrino Yukawa couplings.

To obtain  predictions for the LFV observables  $B(\mu^- \to e^-\gamma
)$, $B(\tau^- \to e^-\gamma)$, $B(\mu^- \to e^-e^-e^+)$, $B(\tau^- \to
e^-e^-e^+)$  and  $B(\tau^-\to e^-\mu^-\mu^+)$,  we  use the  analytic
expressions (4.9) and (4.10)  of~\cite{IP}, along with the formfactors
given in~(\ref{Fgamma})--(\ref{Fbox}).  The predicted rate $R_{\mu e}$
for coherent $\mu  \to e$ conversion in a  nucleus with atomic numbers
$(N,Z)$ may be calculated by
\begin{equation}
  \label{Bmue} 
R_{\mu     e}\      =\     \frac{\alpha^3     \alpha^4_w     m^5_\mu\,
  |F(-m^2_\mu)|^2}{16\, \pi^2 M^4_W \Gamma_{\rm capt}}\ \frac{Z^4_{\rm
    eff}}{Z}\ |Q_W|^2\; ,
\end{equation}
where  $\alpha = e^2/(4\pi)$,  $Z_{\rm eff}$  is the  effective atomic
number of coherence~\cite{COKFV}, $F(-m^2_\mu)$ is a nucleus-dependent
nuclear form  factor~\cite{KKO} and  $\Gamma_{\rm capt}$ is  the total
muon capture  rate. In addition, $Q_W =  V_u (2Z +N) +  V_d (Z+2N)$ is
the weak matrix element, where
\begin{eqnarray}
  \label{VuVd}
V_u \!\!&=&\!\! -\, \frac{2}{3} s^2_w \Big( F^{\mu e}_\gamma +
G^{\mu e}_\gamma + F^{\mu e}_Z \Big) + \frac{1}{4} 
\Big( F^{\mu e}_Z -  F^{\mu e uu}_{\rm box}\Big)\, ,\nonumber\\
V_d \!\!&=&\!\! \frac{1}{3} s^2_w
\Big( F^{\mu e}_\gamma + G^{\mu e}_\gamma
+ F^{\mu e}_Z \Big) - \frac{1}{4} \Big( F^{\mu e}_Z +
F^{\mu e dd}_{\rm box} \Big)\, ,\quad\
\end{eqnarray}
with  $s_w \equiv  \sin\theta_w$.   The leading  contributions to  the
formfactors  $F^{\mu e  uu}_{\rm box}$  and $F^{\mu  e  dd}_{\rm box}$
pertinent to  the up- and  down-quarks, respectively, are  obtained by
calculating the $W$- and $\widetilde{W}$-mediated box graphs analogous
to Fig.~\ref{f1}. More explicitly, we find
\begin{eqnarray}
  \label{Fboxmue}
(F^{\mu  e uu}_{\rm box})^N \!&=&\! -4\,(F^{\mu  e dd}_{\rm box})^N\ =\ 
4\, {\bf \Omega}_{e\mu}\; ,\nonumber\\
(F^{\mu  e uu}_{\rm box})^{\widetilde{N}} \!&=&\! \frac{2\, M^2_W\,
  \widetilde{m}^2_W}{\widetilde{M}^4_Q}\; {\bf \Omega}_{e\mu}\;,\\
(F^{\mu  e dd}_{\rm box})^{\widetilde{N}}\!& =&\! -\;
\frac{M^2_W}{2\,\widetilde{M}^2_Q}\, {\bf \Omega}_{e\mu}\; ,\nonumber
\end{eqnarray}
where we  assumed that $|\widetilde{m}_W|  \ll \widetilde{M}_Q \approx
M_{\tilde{\nu}}$,   with  $\widetilde{M}_Q$   being   a  common   soft
SUSY-breaking  mass for  the left-handed  up and  down squarks  in the
loop.

\begin{figure}[t]
 \centering
 \includegraphics[clip,width=0.45\textwidth]{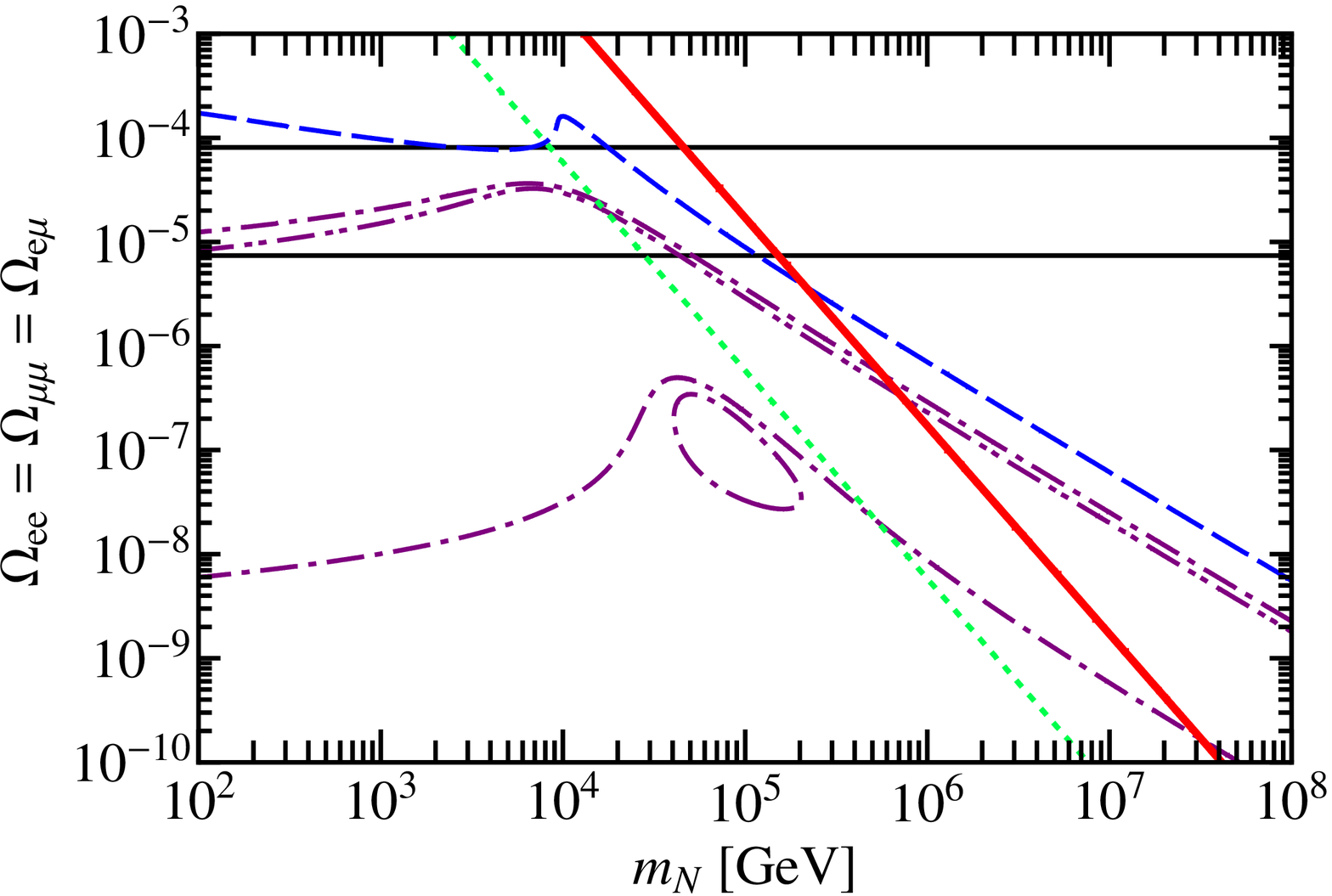}\\[3mm]
 \includegraphics[clip,width=0.45\textwidth]{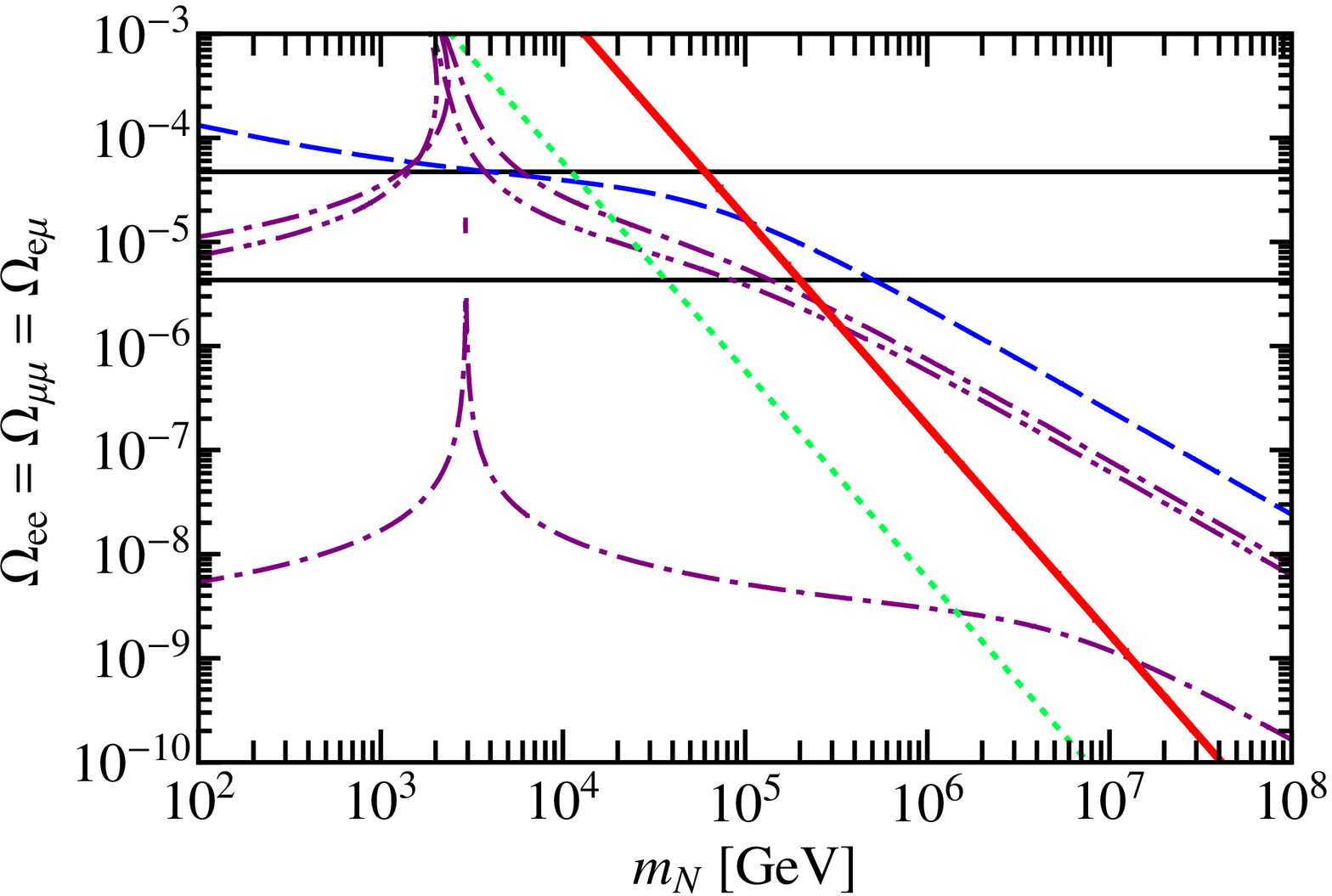}
\caption{Exclusion  contours  of  ${\bf \Omega}_{e\mu}$  versus  $m_N$
  derived  from  experimental   limits  on  $B(\mu^-  \to  e^-\gamma)$
  (solid), $B(\mu^- \to e^-e^-e^+)$ (dashed) and $\mu\to e$ conversion
  in  Titanium (dash-dotted)  and Gold  (dash-double-dotted), assuming
  ${\bf  \Omega}_{ee} = {\bf  \Omega}_{\mu\mu} =  {\bf \Omega}_{e\mu}$
  and ${\bf \Omega}_{\tau\tau} = 0$.   In the lower panel, the quantum
  effects due to $\widetilde{N}_{1,2,3}$  have been ignored. The areas
  that lie  above or within the  contours are excluded;  see the text
  for more details.}\label{f2}
\end{figure}

In our numerical estimates,  we fix $\widetilde{M}_Q = M_{\tilde{\nu}}
=  -\mu =  200$~GeV, $M_{\widetilde{W}}  = 100$~GeV  and  $\tan\beta =
3$~\cite{function}. We first analyze the impact of SLFV on $\mu \to e$
transitions.    We  consider  a   conservative  scenario   with  ${\bf
\Omega}_{ee} = {\bf \Omega}_{\mu\mu}  = {\bf \Omega}_{e\mu}$ and ${\bf
\Omega}_{\tau\tau}  = 0$,  which  yields the  weakest limits  on~${\bf
\Omega}_{e\mu}$.  To  this end, we present  in Fig.~\ref{f2} exclusion
contours for  ${\bf \Omega}_{e\mu}$ versus $m_N$  derived from present
experimental limits and  future sensitivities: $B(\mu^- \to e^-\gamma)
< 1.2\times 10^{-11}$~\cite{PDG} (upper horizontal line), $B(\mu^- \to
e^-\gamma) \sim 10^{-13}$~\cite{MEG} (lower horizontal line), $B(\mu^-
\to e^-e^-e^+) < 10^{-12}$~\cite{PDG}  (dashed line).  We also include
constraints  from  the non-observation  of  $\mu\to  e$ conversion  in
${}^{48}_{22}$Ti  and   ${}^{197}_{\  79}$Au~\cite{Comment1},  $R^{\rm
Ti}_{\mu  e} <  4.3\times 10^{-12}$~\cite{Titanium}  (dash-dotted) and
$R^{\rm     Au}_{\mu     e}     <    7\times     10^{-13}$~\cite{Gold}
(dash-double-dotted),  as  well  as  potential limits  from  a  future
sensitivity    to   $R^{\rm   Ti}_{\mu    e}$   at    the   $10^{-18}$
level~\cite{PRISM} (lower dash-dotted line).  The areas lying above or
within the contours are excluded  by the above considerations.  In the
upper panel  of Fig.~\ref{f2}, the loop effects  from both $N_{1,2,3}$
and $\widetilde{N}_{1,2,3}$  are considered, whereas in  the lower one
only  these from  $N_{1,2,3}$ are  taken into  account.   Finally, the
diagonal dotted  line indicates the regime where  terms $\propto ({\bf
\Omega}_{l'l})^2$ dominate the LFV  observables, whilst the area above
the  diagonal solid  line  represents a  non-perturbative regime  with
${\rm Tr}\,  ({\bf h}^\dagger_\nu {\bf  h}_\nu) > 4\pi$,  which limits
the validity of our predictions.

Figure~\ref{f2}  also  shows  the  importance  of  synergy  among  the
different  LFV  experiments.   In  particular,  we  find  an  area  of
cancellation  in the  predicted value  for $R_{\mu  e}$ for  $m_N \sim
3$~TeV in the non-SUSY case (lower  panel).  This area is covered to a
good extent  by the present limits  from the $\mu  \to eee$ experiment
and by  the current and  future~\cite{MEG} exclusion limits  on $B(\mu
\to e\gamma)$.  Most interestingly, the projected PRISM experiment for
$R^{\rm   Ti}_{\mu   e}   \sim   10^{-18}$~\cite{PRISM}   will   reach
sensitivities to the unprecedented  level of ${\bf \Omega}_{e\mu} \sim
10^{-10}$ and $m_N  \sim 10^8$~GeV.  In the kinematic  regime of large
Yukawa  couplings,  we  see  that  the experiments  for  $\mu  \to  e$
conversion  in nuclei  offer  the highest  sensitivity.   In the  same
regime, we  observe that the  derived bounds on  ${\bf \Omega}_{e\mu}$
and $m_N$  are much stricter in  the SUSY rather than  in the non-SUSY
case (lower  panel).  The reason is  that in this  large $m_N$ domain,
$(F^{l'l}_Z)^{\widetilde{N}}$  prevails over $(F^{l'l}_Z)^N$  and adds
constructively  to  the  dominant  contribution  $(F^{\mu  e  uu}_{\rm
box})^N$.

\begin{figure}[t]
 \centering
  \includegraphics[clip,width=0.43\textwidth]{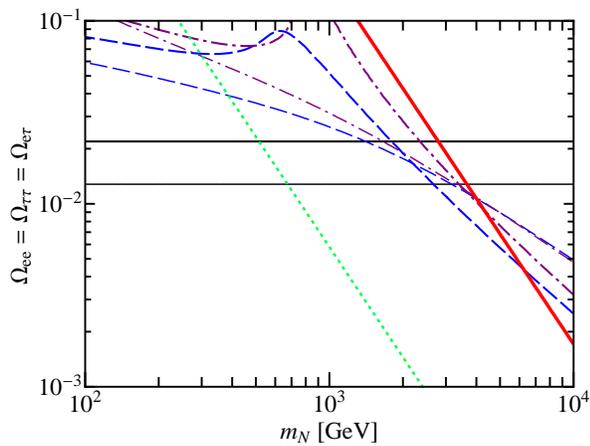} 
\caption{Exclusion  contours of  ${\bf  \Omega}_{e\tau}$ versus  $m_N$
  derived  from present  experimental  upper limits  on $B(\tau^-  \to
  e^-\gamma)$   (solid),  $B(\tau^-   \to  e^-e^-e^+)$   (dashed)  and
  $B(\tau^-\to  e^-\mu^-\mu^+)$  (dash-dotted),  assuming  that  ${\bf
  \Omega}_{ee} =  {\bf \Omega}_{\tau\tau} =  {\bf \Omega}_{e\tau}$ and
  ${\bf  \Omega}_{\mu\mu}  =  0$.   More  details  are  given  in  the
  text.}\label{f3}
\end{figure}

We  now turn  our attention  to $\tau$  LFV, analyzing  a conservative
scenario  with ${\bf  \Omega}_{ee}  = {\bf  \Omega}_{\tau\tau} =  {\bf
\Omega}_{e\tau}$ and ${\bf  \Omega}_{\mu\mu} = 0$~\cite{Comment2}.  In
Fig.~\ref{f3} we display  exclusion contours of ${\bf \Omega}_{e\tau}$
versus $m_N$,  using the present  experimental upper limits~\cite{PDG}
on  $B(\tau^- \to  e^-\gamma)  < 1.1  \times  10^{-7}$ (solid  lines),
$B(\tau^-  \to  e^-e^-e^+)  <  3.6\times 10^{-8}$~(dashed  lines)  and
$B(\tau^-   \to  e^-\mu^-\mu^+)   <   3.7\times  10^{-8}$~(dash-dotted
lines). The thick  lines show exclusion contours of  SLFV, whereas the
thin lines of  the same pattern are the  corresponding contours in the
non-SUSY case. As in Fig.~\ref{f2}, the diagonal dotted line indicates
the regime  where large Yukawa coupling effects  dominate, whereas the
diagonal solid line places the boundary for non-perturbative dynamics.
We  see that  the current  bound  on $B(\tau  \to e\gamma  )$ is  less
sensitive to SLFV,  due to the screening coming  from the right-handed
sneutrinos     in      the     loop.      Moreover,      given     the
constraints~\cite{OmegaDiag},  a  positive  signal for  $B(\tau^-  \to
e^-e^-e^+)$ close  to the present  upper bound will signify  that SLFV
originates   from    rather   large   Yukawa    couplings   and   $m_N
\stackrel{>}{{}_\sim} 3$~TeV.

In summary,  we have shown  that low-mass right-handed  sneutrinos can
sizeably contribute to  observables of LFV.  Thanks to  SUSY, they can
significantly screen  the respective effect of the  heavy neutrinos on
the photonic $\mu$  and $\tau$ decays.  Hence SLFV  can be probed more
effectively  in  present  and   future  experiments  of  $\mu  \to  e$
conversion in nuclei.  The 3-body  decay observables, such as $\mu \to
eee$ and $\tau \to eee$, provide valuable complementary information on
LFV.   In particular, the  former eliminates  a kinematic  region that
remains  unprobed in  the  non-SUSY  case by  $\mu  \to e$  conversion
experiments.  Therefore, plans for  potentially upgrading the $\mu \to
eee$ experiment should  be followed with the same  degree of vigour in
the  community.   In  the  same  vein, the  implications  of  SLFV  in
semileptonic  $\tau$ decays  and in  processes involving  $K$  and $B$
mesons should  also be explored. We  plan to report  progress on these
issues in the near future.


\begin{thebibliography}{99}
\bibitem{OmegaDiag} For the diagonal entries, we conservatively assume
  that $\Omega_{ee,\mu\mu,\tau\tau} \stackrel{<}{{}_\sim}
  10^{-2}$~\cite{BK}.
\bibitem{Gauge}  We  note   that  unlike  the  off-shell  formfactors,
  physical amplitudes do not depend on the choice of the gauge.
\bibitem{CommentSUSY} We do not consider here the infra-red singular SUSY
  limit, without electroweak-symmetry breaking for $\mu \neq 0$.
\bibitem{function} For  this choice of parameter  values, the residual
  loop function $f$ is given by $f \approx M^2_W/(3\widetilde{m}^2_2)$.
\bibitem{Comment1}  To get  predictions for  $R_{\mu e}$,  we  use the
values  for  $Z_{\rm eff}$,  $|F(-m^2_\mu)|$  and $\Gamma_{\rm  capt}$
given in~\cite{KKO}.
\end{thebibliography}
\end{document}